# A Decentralised Multi-Agent Reinforcement Learning Approach for the Same-Day Delivery Problem

**Elvin Ngu**
Centre for Transport Studies, Department of Civil and Environmental Engineering, Imperial College London, London SW7 2AZ, United Kingdom
Email: elvin.ngu17@imperial.ac.uk

**Leandro Parada, Corresponding Author**
Centre for Transport Studies, Department of Civil and Environmental Engineering, Imperial College London, London SW7 2AZ, United Kingdom
Email: l.parada-pradenas20@imperial.ac.uk

**Jose Javier Escribano Macias, Ph.D.**
Centre for Transport Studies, Department of Civil and Environmental Engineering, Imperial College London, London SW7 2AZ, United Kingdom
Email: jose.escribano-macias11@imperial.ac.uk

**Panagiotis Angeloudis, Ph.D.**
Centre for Transport Studies, Department of Civil and Environmental Engineering, Imperial College London, London SW7 2AZ, United Kingdom
Email: p.angeloudis@imperial.ac.uk

ABSTRACT
Same-Day Delivery services are becoming increasingly popular in recent years. These have been usually modelled by previous studies as a certain class of Dynamic Vehicle Routing Problem (DVRP) where goods must be delivered from a depot to a set of customers in the same day that the orders were placed. Adaptive exact solution methods for DVRPs can become intractable even for small problem instances. In this paper, we formulate the SDDP as a Markov Decision Process (MDP) and solve it using a parameter-sharing Deep Q-Network, which corresponds to a decentralised Multi-Agent Reinforcement Learning (MARL) approach. For this, we create a multi-agent grid-based SDD environment, consisting of multiple vehicles, a central depot and dynamic order generation. In addition, we introduce zone-specific order generation and reward probabilities. We compare the performance of our proposed MARL approach against a Mixed Inter Programming (MIP) solution. Results show that our proposed MARL framework performs on par with MIP-based policy when the number of orders is relatively low. For problem instances with higher order arrival rates, computational results show that the MARL approach underperforms the MIP by up to 30%. The performance gap between both methods becomes smaller when zone-specific parameters are employed. The gap is reduced from 30% to 3% for a 5x5 grid scenario with 30 orders. Execution time results indicate that the MARL approach is, on average, 65 times faster than the MIP-based policy, and therefore may be more advantageous for real-time control, at least for small-sized instances.

**Keywords:** Same-Day Delivery, Multi-Agent Reinforcement Learning, Deep Q-Networks, Dynamic Vehicle Routing Problem

**INTRODUCTION**
Same-Day Delivery (SDD) services as used by online retailers seek to deliver goods to customers in less than 24 hours after the order is placed. SDD services play an integral part in e-commerce by allowing online retailers to offer the immediacy of in-person shopping to their consumers. In fact, SDD market size is predicted to grow from $4.7 billion to $9.6 billion in 2022 in the United States alone (1). This increase in demand is likely to have accelerated even more due to consumers massively adopting online shopping during the COVID-19 pandemic.

However, this upward trend in SDD services has led to new challenges related to sustainability, costs, and aging workforce. According to (2), an increase in delivery vans entering city centres negatively impacts the environment, human health and road safety. Labour costs associated with traditional van deliveries are high and an aging workforce in many industrialised countries exacerbates the problem of labour shortage for such a physically demanding job (3). Furthermore, same-day delivery may reduce efficiencies by diminishing the chance for consolidated deliveries with a full load. This worsens the impact of last mile delivery on the environment. Hence, as the SDD market continues to grow, it becomes increasingly important to optimise its operations.

Since SDD services have only recently been widely adopted, the literature in the context of the Dynamic Vehicle Routing Problem (DVRP) is relatively scarce. A few of the recent approaches in the literature include a sample-scenario planning approach, a dynamic dispatch waves approach and mixed integer programming. These studies will be further analysed in our Literature Review.

Exact solution methods for DVRPs rarely scale past a few vehicles (4). Reinforcement Learning (RL), on the other hand, has arisen as a powerful method that can potentially scale to thousands of vehicles and orders. Recently, the combination of Deep Neural Networks (DNN) with RL has shown to achieve superior performance against MIP solvers in similar fleet management problems (6). In addition, DNNs allow for offline training and real-time execution, which is crucial for the SDDP.

To the best of the authors' knowledge, only (5) have implemented Reinforcement Learning (RL) to solve an SDD problem (SDDP) involving heterogeneous fleets of drones and vehicles. However, the RL developed only solves the assignment part of the problem of whether an order is assigned to a drone or a vehicle. Other decisions such as route planning and pre-emptive depot return are solved using separate heuristics and algorithms which may lead to sub-optimal policies. Outside the SDD literature, RL has been used to solve a similar on-demand pick-up and delivery (ODPD) problem. In (6), authors showed that RL outperforms a Mixed Integer Programming solution approach to the ODPD problem, for a single vehicle scenario.

This paper makes an important contribution to the SDD literature by exploring the implementation of a state-of-the-art decentralised multi-agent Deep RL (MARL) approach. More specifically, we adopt a multi-agent version of the well-known Deep Q-Network method, called Parameter-Sharing DQN. The main advantage of this method over MIP-based solvers is the real-time execution (after offline training), which is particularly important in SDD to meet customers' expectation of immediate response. In addition, the MARL formulation gives extra flexibility by allowing agents to determine policies which can strategically wait at the depot, pre-emptively return to depot, reject services or execute unconventional routes in anticipation of future orders. These flexibilities have been shown to significantly improve performance (7-9). Nevertheless, one of the biggest challenges of RL for VRP-related problems is the large action space, which can limit algorithm scalability.

In short, this paper aims to further explore and fill in the gaps in SDD literature with the following step-by-step objectives:
- To model the SDDP as a Markov Decision Process (MDP) with an action space that allows for different agent strategies, such as the ones described above.
- To develop a virtual environment based on the MDP model as a simulator for the MARL framework.
- To implement a state-of-the-art Mixed Integer Programming (MIP) solution approach for the SDDP as a benchmark for the MARL approach.
- To assess the performance and scalability of the proposed MARL approach.

The rest of the paper is structured as follows. First, we discuss the literature related to the SDDP and the RL solution approach. Then, a general description of the SDD problem is given, along with a general MDP model formulation, followed by a presentation of the RL solution approach, or more specifically the parameter-sharing Deep Q-network (DQN) algorithm. The results section describes the set-up of the experiments, the implementation of the DQN, the benchmark algorithm and lastly, the results of the experiments. Finally, the paper is concluded and the potential directions for future work are discussed.

## SAME-DAY DELIVERY AND VEHICLE ROUTING PROBLEMS

### Same-day Delivery Problem: General Background

The SDDP involves the delivery of goods from a depot to customers, with customer requests coming in over the course of a day. The goods must be picked up from a depot before delivery can occur and must be delivered within the same day. SDD-specific literature is relatively scarce as it is very recent, although it is becoming an increasingly popular service in the real world.

In (7), the authors considered the SDD problem for a fleet of vehicles and approached the problem by considering multiple scenarios for future requests to make decisions on whether to accept a request or not. Similarly, (8) solved the same problem with the same sample-scenario planning approach but differs in that it allows vehicles to strategically wait at the depot in anticipation of future requests that can fit into the current planned route. This additional flexibility is particularly useful when orders are heterogeneously spaced in time. In contrast to waiting, (9) investigated a strategy that allows vehicles to pre-emptively return to depot before serving all customers. They found that this strategy increases the number of customers served per workday using a combination of routing heuristic together with approximate dynamic programming methods.

Several studies have approached the SDD problem by formulating it as a dynamic dispatch waves problem (DDWP) (10-12). The DDWP only allows starting of routes at certain dispatch epochs, in this case, every hour. The first two mentioned papers consider only a single delivery vehicle and implement an a-priori policy which applies the rollout algorithm to decide when to leave the depot and which customers to serve. On the other hand, (12) formulates a MDP model. For large spaces, they solve the problem using an integer linear program together with linear value function approximation.

In (13), the authors considered the SDD problem with a focus on local small retail stores as depots with crowd-shippers as the delivery vehicle. Assignment decisions are made based on a mixed integer linear programming (MILP) model with a rolling horizon structure to consider future requests. (14) explored an on-demand meal delivery with drones. Similarly, a MILP model is also devised to represent the problem while using heuristics to solve dynamic part of the problem.

All previous papers explore only homogenous fleets, but there are a few recent papers dedicated to studying the SDD problem with heterogenous fleets (SDDPHF). (14) are the first to study the incorporation of drones into a fleet of vehicles in the context of SDD, applying a parametric Policy Function Approximation (PFA) approach. (6) improved upon this by using a deep Q-learning approach, which demonstrated superior performance when compared to PFA. The authors attributed this improvement to DQN's ability to incorporate more information into its decision-making process such as availability of resources and demands. Furthermore, both papers used a technique novel to the SDD order assignment problem—the minimum cost insertion heuristic.

### Multi-agent Reinforcement Learning in VRP

Multi-agent reinforcement learning (MARL) refers to multiple learnable agents, that take actions and receive back rewards from the same environment (16). Agents in the same environment can be trained independently or cooperatively, where the latter allows communication between agents. In (16), authors showed that cooperative Q-learning can significantly outperform independent Q-learning in many distinct settings if it can be used efficiently. Cooperative training benefits from the additional observations from other agents and results in a faster learning speed, but at a cost of communication. (18) applied deep Q-networks (DQN) in a MARL setting which showed the potential of DQN as a tool in decentralised learning of multiagent systems.

In the context of SDD, decentralised multi-agent RL has yet to be explored, but in the broader literature of DVRP, there are more relevant works exploring decentralised multi-agent RL. (6) implemented DQN to optimise the route of a single delivery driver in a stochastic and dynamic and environment. Their work has shown that the RL agent can consistently outperform a state-of-the-art MIP. However, the training time was roughly three days for a single vehicle in an 8x8 grid map with 10 maximum orders. This may lead to unacceptable training time for scenarios with larger fleets. (19) showed that careful design of the simulator can allow scalability to large-scale fleet management systems. Contextual deep Q-learning and contextual multi-agent actor critic algorithms are used to achieve explicit coordination between agents, which also outperformed state-of-the-art approaches in empirical studies. In contrast to the former paper, the latter successfully scaled up to a 504 hexagonal grid with thousands of orders. Nevertheless, it is still important to note that the problem solved by these two authors are significantly different. The latter studied the fleet repositioning problem which has a smaller state-action space when compared the former, which studied the on-demand pick-up and delivery problem with a much larger state-action space.

In summary, DQN has been shown to be useful in decentralised learning of multi-agent systems. However, scalability to city-sized instances has only been attempted for problems where the RL agents' actions are limited. For flexible policies in which agents have flexible action choices, only a single vehicle scenario has been tested (6). This paper will first test the solution approach by (6) on the SDDP for a single agent and then attempt to scale it up by utilizing a decentralised multi-agent system, the parameter-sharing DQN.

The solution proposed in this paper also aims to take advantage of many of the strategies employed by the SDD literatures discussed earlier, which has yet to be done, possibly due to the complexity in implementing multiple flexibilities using only heuristics. For example, in (8) vehicles are allowed to strategically wait at depot but do not allow a pre-emptive depot return. RL methods can easily implement these strategies by designing a flexible action space such that it can anticipate future requests, strategically wait at the depot as well as pre-emptively return to depot. However, it is worth noting that these added flexibilities do not guarantee that the RL agent can learn to exploit all of them in an effective manner.

## METHODOLOGY

### Problem Description

The problem involves a fleet of vehicles to deliver parcels from a depot to customer whose requests are stochastic and dynamic across the period of a day. Although the probability distribution of where and when the customer will appear is known, the actual location and time of customer request are unknown until revealed.

When a request comes in, each vehicle in the environment is required to accept or reject the request. The vehicle must make this decision within the next time step to simulate real-world SDD service providers, which give immediate feedback to the customers on whether a SDD service is available.

Upon order acceptance, the parcel must be delivered within a fixed deadline, and missing a deadline will result in a penalty. Penalties will be given to any accepted but missed orders because this will mean customer expectation is not being met, resulting in low customer satisfaction. If more than one vehicle accepts the order at the same time step, the order will be assigned to the agent with the minimum insertion cost. This assignment is made irreversible since the process of packaging and loading onto a specific vehicle would have begun.

Among the assigned orders, the vehicle can also decide the route plan for which orders to serve first. In addition, the vehicles can choose to wait strategically at the depot in anticipation of future requests. While en route, agents are also allowed to pre-emptively return to depot to consolidate the delivery route. If an order is rejected by all vehicles, no penalty or reward will be given as it is assumed that the order is simply assigned to another delivery service such as next day delivery.

**Model Preparation**

The SDDP is modelled as a Markov Decision Process (MDP). The fleet of $m$ identical vehicles is denoted as $\mathcal{F} = \{v_1, v_2 \ldots, v_m\}$ with their respective locations, $L = \{l(v_1), l(v_2) \ldots, l(v_m)\}$ initialised at the depot location, denoted as $l(d)$ at the beginning of episode, time $t_0$. The subscript in set $\mathcal{F}$ is referred to as the vehicle ID denoted as $i$. Throughout the episode, customer orders come in until a fixed cut-off time $t_c$. The set of $n$ customers is denoted as $\mathbb{C} = \{c_1, c_2 \ldots, c_n\}$. The time of order and location of customer orders is denoted by $t(c_j)$ and $l(c_j)$ respectively, where $j$ represents the customer ID. The time window to deliver the order is between $t(c_n)$ and $t(c_n) + \delta$, where $\delta$ is a fixed time length between the order time and order deadline. If the fixed order deadline $t(c_n) + \delta$ is after the end of episode time $t_e$, then the order deadline is set to $t_e$. All vehicles are required to return by time $t_e$. The time taken to pick up and load a package onto the vehicle is $t_p$ whereas the time taken to deliver an order upon reaching customer location is $t_d$.

**Markov Decision Process Model Formulation**

The SDD problem was modelled as in (9, 15). There are 5 main components to an MDP – decision point, state space, action space, rewards, and transition. The decision point is defined as the time at which a decision is made. For this problem, a decision is required at every time step. Hence, the time step representing the $k^{th}$ decision point is denoted as $t_k$.

The state contains all the information needed to make the decision at a particular decision point. The state at time, $t_k$ is denoted as state $S_k$. For this problem, the state contains the following information:
- $t_k$: time of $k^{th}$ decision point *i.e.* the current time step
- $L = \{l(v_1), l(v_2) \ldots, l(v_m)\}$: vehicle locations
- $\Omega = \{l(c_1), l(c_2) \ldots, l(c_n)\}$: customer order locations
- $\Phi = \{o(c_1), o(c_2) \ldots, o(c_n)\}$: customer order statuses
- $T = \{t(c_1), t(c_2) \ldots, t(c_n)\}$: customer order time of requests

The order status has the following values:
- $o(c_j) = -1$, if the order of customer $c_j$ is inactive.
- $o(c_j) = 0$, if the order of customer $c_j$ is open and not yet accepted.
- $o(c_j) = 1$, if the order of customer $c_j$ has been accepted and assigned to another vehicle in the environment.
- $o(c_j) = 2$, if the order of customer $c_j$ has been accepted by the vehicle observing this state space but has yet to be picked up by the vehicle at the depot.
- $o(c_j) = 3$, if the order of customer $c_j$ has been accepted by the vehicle observing this state space and has been picked up and loaded onto the vehicle.
- $o(c_j) = 4$, if the order of customer $c_j$ has been successfully delivered to the customer before deadline.

The state is mathematically defined as:

$$S_t = (t, L, \Omega, \Phi, T) \tag{1}$$

At each decision point, an action within the action space is to be selected. Each vehicle in the set $\mathcal{F}$ is required to select their own action, $x_k$ at all decision points, $t_k$. The number of actions within an action space is dependent on the episode's setting, more specifically the setting that determines the maximum number of orders that can be simultaneously active during a time step. Nevertheless, the action space is always fixed throughout an episode with a particular setting. At decision point $t_k$, the vehicle can decide between the following actions.

$x_k$ consists of the following possible actions:
- $x_k = 0$: wait and do nothing
- $x_k = 1$: accept customer request
- $x_k = 2$: move a step towards the depot. Nothing happens if already at depot.
- $x_k = j + 2$: move a step towards the customer location, $l(c_j)$.
- $x_k = n + 2$: move a step towards the customer location, $l(c_n)$.

Hence, the number of available actions is $n + 2$. This action space is particularly large especially during a large-scale simulation when the number of orders that needs to be simultaneously active becomes very large. In a decentralised scenario, it is possible for more than one vehicle to compete to accept the same order. As described earlier, a minimum cost insertion algorithm is used to resolve this conflict. The minimum cost insertion algorithm calculates the increase in distance when an extra order is added into a current route. If the vehicle is still in depot, all assigned orders are included as the current route. The difference between the total distance of the new route with the newly accepted order and the total distance of the current route is calculated. The order is then assigned to the vehicle that has the minimum cost increase due to the new order insertion. If the vehicle is not at the depot and is delivering orders in a route, then the orders that are already packed into the van is not included in this minimum cost insertion calculation. Only the orders assigned to the vehicle but has yet to be picked up at the depot is included in the route distance calculation.

The reward for a state-action pair is denoted as $R(S_k, x_k)$. Let $r$ be the total reward for successfully delivering an order. If an order is accepted, a third of the reward $r/3$ is given. Once the delivery is successfully, the remaining two-thirds of the reward $2r/3$ is given. No reward or penalty is given for rejected customers. A negative reward or penalty is given for any invalid actions. Invalid actions include accepting order when no order is open or attempting to deliver an inactive order.
- $R(S_k, x_k) = r/3$, for successfully accepting an order.
- $R(S_k, x_k) = 2r/3$, for successfully delivering an order before deadline.
- $R(S_k, x_k) = -\pi_1$, for invalid actions.
- $R(S_k, x_k) = -\pi_2$, for missing deadline of assigned orders.
- $R(S_k, x_k) = -\pi_3$, for failing to return to depot at the end of episode.
- $R(S_k, x_k) = 0$, otherwise

transitions to the post-decision state, $S_{k,p}$. The post-decision state will update deterministic changes, as a result of the action taken by the fleet. From the post decision state, $S_{k,p}$, the environment transitions to the next pre-decision state, $S_{k+1}$. During this transition, exogenous information are revealed which is independent of the actions taken by the vehicles. In this case, the exogenous information is the generation of orders and customer's location. The process terminates $t_k = t_e$.

**REINFORCEMENT LEARNING SOLUTION APPROACH**

The Deep Q-Network (DQN) is a modified version of the simpler Q-learning algorithm, which is a RL method that learns the value of a state-action pair, known as Q-values. Each Q-value is an estimate of the expected future reward for taking an action in any given state. A table containing the Q-values of all possible state-action pair is known as a Q-table.

However, it is virtually impossible to exhaustively explore all the possible Q-values for tabulation in complex environments with multiple agents. Hence, a deep neural network is used to approximate Q-values, using a given set of features from the state space as inputs. This is known as the DQN and it is based on the following loss function:

$$L(\theta) = \mathbb{E}_{s_t, a_t, s_{t+1}}[r_{t+1} + \gamma \max_a Q_{\theta'}(s_{t+1}, x) - Q_\theta(s_t, x_t)] \qquad (2)$$

,where $Q_\theta$ and $Q_{\theta'}$ are both neural networks with sets of parameters $\theta$ and $\theta'$, respectively. The $Q_{\theta'}$ corresponds to the target network and is not updated every through gradient descent, but simply copied from $Q_\theta$ every certain number of episodes.

We use a parameter-sharing DQN framework, i.e., each agent can learn an independent policy, but all agents share the parameters of the network. This makes training more scalable than the fully decentralized approach, because the number of trainable parameters does not depend on the number of agents.

**EXPERIMENTS**

The experiments were run on a grid world environment, where agents can only move north, south, east or west. Each episode consists of 144 time steps ($t_e = 144$), to reflect a working 12-hour delivery day with one time step representing 5 minutes in the real world. The vehicle speed is set to one grid per unit time. The time required to deliver, pickup and accept orders are set as $t_d = t_p = t_a = 1$.

The default depot location is set at the centre of grid. If the grid map is an even number and there are four centre locations, then the depot will be set to any one of the center grids. For example, the depot location can be at (5,5) or (6,6) in a 10x10 grid map. A representation of the grid world is shown in the Figure 1.

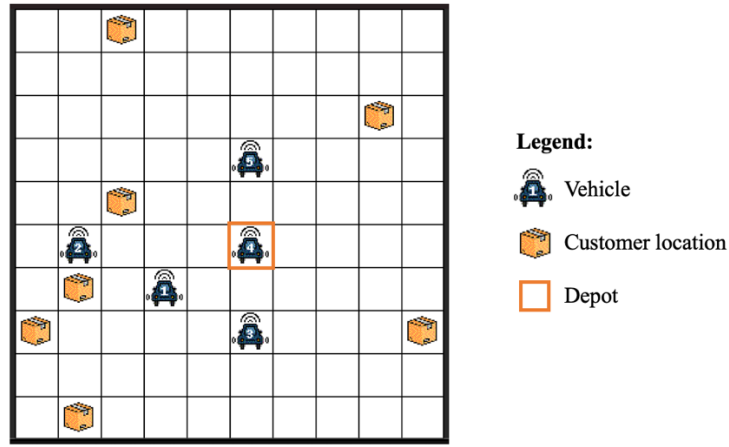

**Figure 1 Illustration of a 10x10 gridworld environment with 5 vehicles.**

The deadline of the order, $\delta$ is set to 48 units time (4 hours in the real world) from the time a customer requests an order, $t(c_n)$. The penalties for invalid action, missed order and failure to return to depot by end of episode is set as $\pi_1 = \pi_2 = \pi_3 = 5$ and the reward for a successful order is set as $r = 3$. Lastly, the number of orders generated in the environment is stochastic as it is set according to the expected number of orders given by the user. At every time step, there is a certain probability that an order will appear such that by the end of the episode, the total number of orders will be roughly equal to the given expected number of orders. The grid map size, number of vehicles and expected number of orders are three of the settings that are varied across experiments which will be specified accordingly.

The same DQN architecture is used for all experiments in this paper. A neural network architecture consists of three hidden layers, consisting of 256, 256 and 128 nodes, respectively. The input layer receives a set of features from the environment, i.e., the state observation described in the MDP formulation. The number of nodes in the output layer corresponds to the number of possible actions in the environment.

A flowchart of each step is presented in figure 2. The full state is retrieved from the environment and passed to the Parameter-sharing DQN, which then outputs the individual Q values for each vehicle. Based on the Q values, the policy outputs the action, which is then passed to the environment to perform vehicle movements.

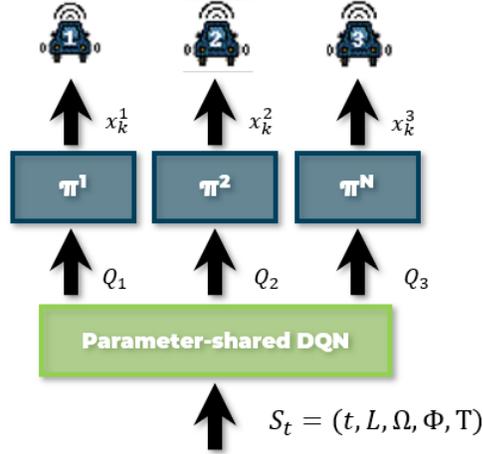

**Figure 2 Flowchart of a step of the proposed solution algorithm.**

**Baseline Algorithm**
    The proposed DQN approach is compared against a baseline model consisting of a deterministic mixed integer programming (MIP) solution approach. (6) also used MIP as a baseline model for the dynamic and stochastic pick-up and delivery problem.
    Whenever a new order arrives, a solution from the deterministic MIP is obtained for the current available orders in the environment. The MIP can either accept or reject the order and will execute a route plan such that it can serve all the currently assigned orders in the shortest possible time.
    A near-optimal solution instead of the optimal solution can be accepted if the runtime for the MIP exceeds the set maximum optimisation time. This is because finding a solution to the MIP can become computationally intractable for more complex scenarios. However, the maximum optimisation time is set such that this occurs in less than 10% of tested episodes.
    It is worth noting that it is difficult to directly compare the performance of algorithms from different papers in the literature review section even though they all solve seemingly similar SDDP. This is due to differences in different papers' set-up and definition of the SDDP, and besides, it is difficult to accurately reproduce the exact problem formulation from another paper because problem descriptions can often possess some ambiguities (9). Hence, the solution approaches from other papers will need to be reproduced to fairly compare the different methods. This is one of the main reason the MIP model is used as a benchmark solution instead of directly comparing results to other SDD papers. Comparison of the DQN with anticipatory models described earlier are left for future works.

The mathematical model for the MIP is now presented:

Sets:
V: Current vehicle location, V = {0}
P: Pickup location (depot location, associated with orders that are not in transit)
D: Delivery locations representing all orders that are not in transit
A: Delivery locations representing the orders that are accepted by driver, but not in transit
T: Delivery locations representing orders that are accepted by driver, but in transit
R: Return location (depot location, used for final return)
N: Set of all nodes/locations in the graph, $N = V \cup P \cup D \cup A \cup T \cup R$

E: Set of all edges, $E = \{(i,j), \forall i,j \in N\}$

Decision variables:
$x_{ij}$: Binary variable, 1 if vehicle uses the arc from node $i$ to $j$, 0 otherwise; $i,j \in N$
$y_i$: Binary variable, 1 if the order $i$ is accepted, 0 otherwise; $i \in D$
$B_i$: Auxillary variable to track the time as of node $i$; $i \in N$

Parameters:
n: number of orders available to pick up, n = |D|
cij: Symmetric Manhattan distance matrix between node i and j; $(i,j) \in E$
li: Remaining time to deliver order i, $i \in D \cup T$
m: Travel cost per mile
ri: Reward for orders associated with deliveries that are not in transit, D
M: A big real number
t: time to travel one mile
d: A constant service time spent on accept, pick up and drop off

Model:
$$\max \sum_{i \in D} r_i y_i - m \sum_{(i,j) \in E} c_{ij} x_{ij} \tag{3}$$

Subject to:
$$\sum_{i \in V} \sum_{j \in N} x_{ij} = 1 \tag{4}$$
$$\sum_{i \in P} \sum_{j \in N} x_{ij} = 1 \tag{5}$$
$$\sum_{j \in N} x_{ij} = y_i \quad \forall i \in D \tag{6}$$
$$\sum_{j \in N} x_{ij} = 1 \quad \forall i \in T \tag{7}$$
$$\sum_{i \in N \setminus R} \sum_{j \in R} x_{ij} = 1 \tag{8}$$
$$\sum_{j \in N \setminus R} x_{ji} - \sum_{j \in N} x_{ij} = 0 \quad \forall i \in P \cup D \cup T \tag{9}$$
$$y_i = 1 \quad \forall i \in A \tag{10}$$
$$B_i + d + c_{ij} t - M(1 - x_{ij}) \le B_j \quad \forall i,j \in N \tag{11}$$
$$B_i + c_{ij} t - M(1 - y_j) \le B_j \quad \forall i \in P, j \in D \tag{12}$$
$$d \sum_{i \in D \setminus A} y_i = B_0 \tag{13}$$
$$B_i \le l_i \tag{14}$$
$$x_{ij}, y_i \in \{0,1\} \quad \forall i,j \in N \tag{15}$$

    Constraints 4 to 7 restrict the flow of vehicles. Constraint 4 ensures the vehicle leaves its current location only once. Constraint 5 ensures vehicle leaves the depot only once for pick up. Constraints 6 and 7 ensure vehicle leaves the delivery destinations only once. Lastly, constraint 8 ensures the vehicle return to the depot. Constraint 9 ties everything together by enforcing a zero net flow through the nodes, hence ensuring that set P, D and T are visited once and only once.

    Constraint 10 ensures that all previously accepted orders are included in the route. Constraints 11 to 15 are time constraints. Constraints 11 ensures that time window is met. Constraint 12 sets the priority that orders not-in-transit must be picked up at depot before being delivered. These two constraint were originally non-linear, and were both linearised using the big M method (20). Constraint 13 ensures time required to accept and deliver orders are accounted for. Equation 14 ensures order to be delivered before expiry. Lastly, Constraint 15 ensures decision variable x and y are binary. It should be noted that the capacity of the vehicle in this problem formulation is unconstrained. This is reasonable for small order sizes, which results in vehicles only taking a limited amount of orders, but would not be realistic for bigger order sizes.

**Results**

*Ngu, Parada, Escribano Macias, and Angeloudis*

For each set of experiments, the DQN was trained over 150,000 episodes and then tested over 100 episodes, while the MIP method was executed over 100 random episodes. Moreover, due to the stochasticity of the environment, 3 independent training rounds were obtained for the DQN algorithm. The total episodic reward is used as the performance measure for each method. It is worth noting that both methods were assessed on the exact same environment, with the same reward function. For training, a workstation was used with an Intel Core i9-10900X CPU processor and a NVIDIA RTX 3090 GPU.

*Single-agent Scenario*

The proposed parameter-sharing DQN method is first compared to the MIP baseline on a single-agent environment. We experimented different environment parameters, considering two grid map sizes (5x5 and 10x10) and two expected number of orders (5 and 30). These values were chosen to show how the size of grid world and number of orders affect the performance of the MIP. The results are shown in Table 1, where there are two sections referring to a homogenous and a heterogenous scenario. Scenarios marked with an asterisk in Table 1 are trained over larger number of episodes to achieve more consistent results over the three independent runs. The homogenous scenario corresponds to an environment where order generation and rewards are the same across the whole map, whereas in the heterogenous scenario order generation and rewards are given by a random probability distribution. For the heterogenous scenario, the grid map was divided into four zones with relative order probabilities of {0.3, 0.4, 0.2, 0.1}, with each zone respectively generating orders with maximum and minimum rewards of {[12,8], [8,6], [5,3], [3,1]}.

Table 1 shows that the proposed DQN method can achieve similar final rewards as the MIP when the order number is small. However, for the problem instance with high number of orders (30), the MIP significantly outperforms the DQN. This is likely due to the DQN method converging to a policy that takes a sub-optimal route. Furthermore, the lower reward obtained by the DQN is also due to some missed orders and failure to return to the depot by the end of episode. In the heterogenous scenario, the DQN approach still underperformed the MIP, but performed more closely to the MIP benchmark. This is especially true for the smaller 5 by 5 map with 30 orders scenario where the DQN only underperforms MIP by 3% in the heterogenous scenario as compared to 15% in the homogenous scenario. Note that a higher number of episodes was needed to achieve convergence for the scenarios consisting of a 10 by 10 map and 30 orders. This is mainly because the higher environment complexity of having more orders.

**TABLE 1 Comparison of RL performance vs MIP for a single-agent scenario.**

|  | **Problem Instance** | **DQN** | **MIP** | **Difference in performance** |
|---|---|---|---|---|
| **Homogenous** | 5 by 5 map, 5 orders | 14.3 | 14.0 | 2.6 % |
| | 5 by 5 map, 30 orders | 62.1 | 72.9 | -14.8 % |
| | 10 by 10 map, 5 orders | 14.0 | 14.2 | -1.4 % |
| | 10 by 10 map, 30 orders* | 34.8 | 50.1 | -30.6 % |
| **Heterogenous** | 5 by 5 map, 30 orders | 103.7 | 106.9 | -3.0 % |
| | 10 by 10 map, 30 orders* | 52.1 | 70.3 | -25.9 % |

*trained over 250,000 episodes instead of 150,000 episodes.

Figure 3 shows the DQN training curve for the 5 by 5 map with 30 heterogenous orders distribution shown in Table 1. The graph shows that the DQN method has large inter-episode variance, and some of the episodes outperform the MIP benchmark. Furthermore, the DQN method occasionally incurs in large penalties of up to -150, which negatively impacts the average performance. This is likely due to the active exploration of state-action space by the DQN algorithm.

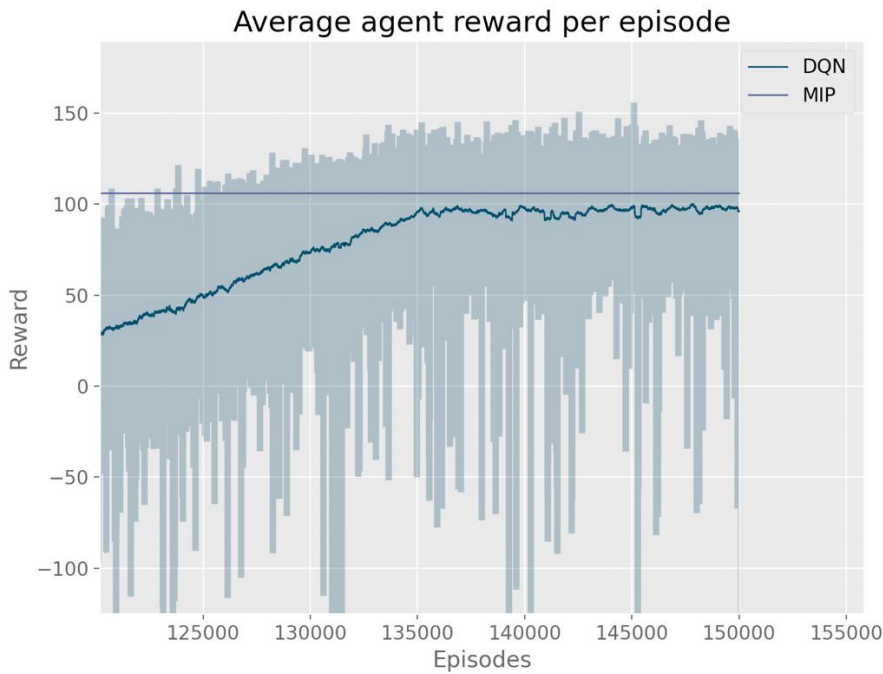

**FIGURE 3 Training curve for the DQN algorithm, considering a 5x5 grid map with 30 heterogeneously distributed orders.**

*Multi-Agent Scenario*

In the following set of experiments, we used a 10x10 grid map with 30 orders, and increased the agent number up to five. Results are summarized in table 2. As shown, positive rewards for all cases were achieved by the MARL approach after 150k training episodes. Furthermore, when the number of agents reached four, the rewards plateau as the agents can virtually deliver all 30 orders. The DQN method performed worse than the MIP-based method for one, two and three agents. However, when the number of agents is increased to four and five, DQN achieved better results than the MIP. This suggests that the DQN approach may be more advantageous than the MIP in environments with higher number of agents.

**TABLE 2 Comparison of RL performance for the multi-agent scenario.**

| Number of Agents | DQN | MIP | Difference in performance |
|---|---|---|---|
| 1* | 34.8 | 50.1 | -30.6% |
| 2 | 37.5 | 69.0 | -31.5% |
| 3 | 64.8 | 74.1 | -12.6% |
| 4 | 78.3 | 71.9 | 9.0% |
| 5 | 78.7 | 73.4 | 7.3% |

*trained over 250,000 episodes instead of 150,000 episodes

*Execution Time Evaluation*
We performed experiments on execution time to assess the efficiency of each method and their suitability for real-time control. Table 3 and Table 4 show the average execution time per episode for each method. Results show that the execution time for the proposed MARL approach is far superior when compared to the MIP-based approach for all scenarios. This is an important advantage since real-time decisions are essential for real world SDD services. The MIP tends to take much longer when the ratio between the number of orders and the grid size is higher. This is because high orders in a small grid will result in more simultaneous active orders, which means that there are more nodes, edges and decision variables that need to be considered by the MIP.

**TABLE 3 Comparison of average execution time between RL and MIP for the single agent scenario.**

| Problem Instance | | DQN execution time (s) | MIP execution time (s) | Execution time ratio (MIP/DQN) |
|---|---|---|---|---|
| Homogenous. Single Agent | 5 by 5 map, 5 orders | 0.15 | 0.50 | 3 |
| | 5 by 5 map, 30 orders | 0.21 | 140.33 | 670 |
| | 10 by 10 map, 5 orders | 0.15 | 1.62 | 10 |
| | 10 by 10 map, 30 orders | 0.22 | 31.47 | 140 |
| Heterogenous. Single Agent | 5 by 5 map, 30 orders | 0.22 | 51.73 | 230 |
| | 10 by 10 map, 30 orders | 0.34 | 39.47 | 110 |

**TABLE 4 Comparison of average execution time between RL and MIP for the multi-agent scenario.**

| Number of Agents | DQN execution time (s) | MIP execution time (s) | Execution time ratio (MIP/DQN) |
|---|---|---|---|
| 1 | 0.22 | 31.47 | 140 |
| 2 | 0.81 | 45.02 | 50 |
| 3 | 1.03 | 16.64 | 15 |
| 4 | 0.71 | 24.18 | 30 |
| 5 | 0.91 | 16.57 | 18 |

**CONCLUSIONS AND FUTURE WORK**

In this paper, we formulated the Same-Day Delivery problem as a Markov Decision Process and solved it using a parameter-sharing DQN, which corresponds to a decentralised multi-agent RL (MARL) approach. An established MIP algorithm was used as a benchmark comparison for the proposed MARL approach. To compare both methods, we designed and implemented an SDD environment consisting of a central depot, multiple delivery vehicles and dynamic order generation. We then experimented on two different scenarios: single-agent and multi-agent.

For the single-agent scenario, we showed that when the order rate is small the DQN approach is, at least, competitive to the MIP-based method. For problem instances with higher order arrival rates, the computational results showed that the MARL approach underperformed the MIP, regardless of the grid size. The reason for this is that the MIP-based solver corresponds to a centralized method, where all decisions are made by a single central unit that optimizes the global objective of the system. In contrast, MARL agents are based on the decentralized learning paradigm, that is, they optimize local decisions using a partial observation of the environment, which may be more realistic in certain real-world applications. When zone-specific order generation and reward probabilities are introduced, the gap between both methods is smaller, which may suggest that the DQN approach can be more advantageous in environments with higher complexity.

For the multi-agent case, we showed that the proposed approach achieves similar performance as the MIP-based method, while being up to 65 times faster during execution, on average. This is a crucial advantage for the SDDP, as decisions must be made in real-time. Moreover, real-world problem instances can scale to hundreds of independent drivers and thousands of order requests within a day. It should be noted that we trained the DQN offline, and used the trained model to develop the real-time control policies. Training times for the DQN approach for all instances took from a few hours up to a day.

The main limitation of our study is that we evaluated the performance of the MARL approach for a small number of agents. We believe that the method should be assessed on a real-sized instance, where stability issues may arise due to agents learning independently. Thus, the next step would be to implement the DQN on a large-scale SDDP environment. Real-world instances will inevitably result in higher environment complexity, and therefore new methods for faster training should be evaluated. Recent efforts in this domain include the use of Curriculum Learning (CL) (21) and the use of Policy Ensembles (22).

It is worth noting that both models can easily incorporate extra constraints, such as vehicle capacity, although adding constraints to the MIP will likely increase the complexity of the problem, and consequently the execution time. In contrast, adding constraints to the MARL formulation will probably not affect execution time, although additional number of episodes may be required to achieve convergence. This is because adding constraints to the MARL formulation is directly related with the environment complexity.

Another potential direction for future research is to explore different combination of input features. For example, some of the information such as distance from orders is implicitly derived from the state space. It is possible that explicit inclusion of such information as input features can improve the DQN's solution. We believe it is also worth to explore the use of different neural network architectures to account for different features of the state space, for instance, using Convolutional and/or Recurrent NNs.

**ACKNOWLEDGMENTS**
This research was partially supported by the Chilean National Agency for Research and Development (ANID) through the "BECAS DOCTORADO EN EXTRANJERO" programme, Grant No. 72210279.

**AUTHOR CONTRIBUTIONS**
The authors confirm contribution to the paper as follows: study conception and design: Angeloudis, Parada and Escribano; model development: Ngu and Parada; results collection and analysis: Ngu; draft manuscript preparation Ngu, Parada and Escribano. All authors reviewed the results and approved the final version of the manuscript.